\documentclass[showkeys,showpacs,prd]{revtex4}
\usepackage{amsmath}
\usepackage{amssymb}
\usepackage{graphicx}
\usepackage{color}

\begin{document}

\title{
Modeling a nonperturbative spinor vacuum \\interacting with a strong gravitational wave
}

\author{
Vladimir Dzhunushaliev$^{1,2}$
\footnote{v.dzhunushaliev@gmail.com}
and Vladimir Folomeev$^{3}$
\footnote{vfolomeev@mail.ru}
}
\affiliation{
$^1$ Department of Theoretical and Nuclear Physics,  Al-Farabi Kazakh National University, Almaty 050040, Kazakhstan \\
$^2$ Institute of Experimental and Theoretical Physics,  Al-Farabi Kazakh National University, Almaty 050040, Kazakhstan \\
%$^3$ Institute for Basic Research, Eurasian National University, %Astana, 010008, Kazakhstan \\
$^3$ Institute of Physicotechnical Problems and Material Science of the NAS of the Kyrgyz Republic, 265 a, Chui Street, Bishkek 720071,  Kyrgyz Republic
}

\begin{abstract}
We consider the propagation of strong gravitational waves interacting with a nonperturbative vacuum of spinor fields. To described the latter, we suggest an approximate model. The corresponding Einstein equation has the form of the Schr\"odinger equation. Its gravitational-wave solution is analogous to the solution of the Schr\"odinger equation for an electron moving in a periodic potential. The general solution for the periodic gravitational waves is found. The analog of the Kronig-Penney model for gravitational waves is considered. It is shown that the suggested gravitational-wave model permits the existence of weak electric charge and current densities concomitant with the gravitational wave. Based on this observation, a possible experimental verification of the model is suggested.
\end{abstract}

\pacs{04.30.-w}
\keywords{strong gravitational waves, nonperturbative vacuum, spinor field}
\date{\today}

\maketitle

\section{Introduction}

An experimental search for  gravitational waves (GW) is one of the most intriguing problems in modern physics.
The discovery of GWs will give us confidence that we are moving in the right direction
%we are well aware
in understanding classical gravity. If we cannot detect GWs, then either
our apparatus is not sensitive enough or something is not taken into account when considering the propagation of GWs.

One of the effects accompanying the  propagation of GWs could be its interaction with a nonperturbative spinor vacuum.
The reason for
such an interaction to occur is that the energy-momentum tensor of a spinor field contains the spin connection, which in turn contains first derivatives
of tetrad components with respect to the coordinates. As a
result, the Einstein equations give the wave equation for a
GW which contains second derivatives of the tetrad components on the left-hand side and their first derivatives on the
right-hand side.

In Ref.~\cite{Dzhunushaliev:2014daa} we have considered
the propagation of a weak GW interacting with the nonperturbative spinor vacuum.
Here we extend those results to the case of a strong GW. In doing so, as in Ref.~\cite{Dzhunushaliev:2014daa},
to model the nonperturbative vacuum of a spinor field,
we will use a phenomenological approach.
Within the framework of this approach, we make some physically reasonable assumptions about expectation
values of the spinor field and its dispersion. This will permit us to reduce the infinite system of differential equations
for all Green's functions of the nonperturbative quantum spinor field to the finite set of equations (for more details, see
Refs.~\cite{Dzhunushaliev:2014daa,Dzhunushaliev:2013nea}).

Following this approach, here we will discuss the solution of the Einstein equations for a strong GW propagating  on the
background of the nonperturbative vacuum of spinor fields,
which is a generalization of the weak, plane gravitational wave of Ref.~\cite{Dzhunushaliev:2014daa}.

\section{Strong GW in a nonperturbative spinor vacuum}

According to the textbook \cite{lanlif}, the metric for a strong GW propagating in one direction is sought in the form
\begin{equation}
	ds^2 = 2 d \xi d \eta + g_{ab}(\eta) dx^a dx^b,
\label{1-10}
\end{equation}
where $x^0 = \xi, x^1 = \eta$ are lightlike  coordinates, and the indices
$a,b$ run over the values $2,3$. It is convenient to introduce the new variables
\begin{eqnarray}
	g_{ab}(\eta) &=& - \chi^2(\eta) \gamma_{ab}(\eta),
\label{1-20} \\	
	\det (\gamma_{ab}) &=& 1 .
\label{1-30}
\end{eqnarray}
The only nonvanishing component of the Einstein tensor $G_{\bar \mu \nu}$ is $G_{\bar 1 \eta}$ ($\bar \mu$ is the tetrad index and $\nu$ is the spacetime index), so that one can obtain the following Einstein equation for the strong GW propagating in a flat vacuum spacetime:
\begin{equation}
	\ddot \chi + \frac{1}{8} \left(
		\dot \gamma_{ac} \gamma^{bc} \dot \gamma_{bd} \gamma^{ab}
	\right) \chi = 0.
\label{1-40}
\end{equation}
Here the dot denotes differentiation with respect to $\eta$;
 $\gamma^{ab}$ is the two-dimensional tensor reciprocal  to
$\gamma_{ab}$. The function $\chi(\eta)$ is an unknown function, and
$\gamma_{ab}$ are arbitrary functions, obeying the constraint \eqref{1-30}.

As pointed out in Ref.~\cite{lanlif},  the presence of the term
$\dot \gamma_{ac} \gamma^{bc} \dot \gamma_{bd} \gamma^{ab}$ has the result that after a finite time interval
%the determinant $\det \chi_{ab}$
$\chi$ becomes zero.
%It happens since it is always negative.
This in turn leads to vanishing of the metric determinant $g$, i.e., a singularity in the metric.
But this singularity is not physically significant since
it is related only to the unsatisfactory nature of the
reference frame, ``spoiled'' by the passing gravitational wave, and can be eliminated by a
suitable coordinate transformation.
%it is related only to the inadequacy of the reference frame furnished by the passing strong GW and can be eliminated by appropriate coordinate transformation.

Our goal here is to consider the propagation of GWs in a nonperturbative spinor vacuum.
We expect that such a physical system has to be considered in nonperturbative language when both a
metric and a spinor field are regarded as quantum quantities and are quantized in a nonperturbative manner.
Then, to describe the interaction between a quantum metric and a quantum spinor field,
we have to write down the Einstein-Dirac operator equations.

%In order to describe such a situation, we have to write down the Einstein-Dirac operator equations
%to describe the interaction between a quantum metric and a quantum spinor field.

Our nonperturbative approach for quantizing nonlinear fields is described in Ref.~\cite{Dzhunushaliev:2014daa}.
Within this method, we have, strictly speaking, to solve the following operator equations:
\begin{eqnarray}
	\hat G_{\bar a \mu} &=& \varkappa \hat T_{\bar a \mu} ,
\label{2-10}\\
	 \hat \gamma^\mu \nabla_\mu \hat \psi - m \hat \psi &=& 0,
\label{2-20}
\end{eqnarray}
where $\hat G_{\mu \nu}$ is the operator of the Einstein tensor;
$\hat T_{\mu \nu}$ is the operator of the energy-momentum tensor;
$\hat\psi$ is the operator of the spinor field; $\bar a$ is the tetrad index; $\mu$ is the coordinate index;
$\nabla_\mu$ is the covariant derivative for the spinor with the appropriate spin connection;
$\varkappa=8\pi G$, $G$ is the gravitational constant. Hereafter we use units where $c=\hbar=1$.

As mentioned in Ref.~\cite{Dzhunushaliev:2014daa},
the solution for
 such a set of operator equations  can be found by writing down and subsequent solving an infinite system of equations
 for all Green's functions for quantum fields.
% and then solving such equations set.
 In practice,  such a procedure can be carried out only approximately.
 This means that we have to use some physical arguments to cut off the above-mentioned \textit{infinite} set of equations, and then solve
 the resulting
% such cutting off
 \textit{finite} set of equations.

Our approach is based on the following assumptions:
\begin{enumerate}
\item[(i)] The metric remains always classical.
\item[(ii)] The spinor field is quantum.
\item[(iii)] The spinor field is decomposed as a product of $q-$ and $c-$numbers.
\item[(iv)] In order to calculate the energy-momentum tensor of the spinor field, we assume some ans\"atze for a 2-point Green's function of $\hat \psi$.
\end{enumerate}

In such an approximation the Einstein-Dirac operator equations \eqref{2-10} and \eqref{2-20} can be written in the following manner~\cite{Dzhunushaliev:2014daa}:
\begin{eqnarray}
	G_{\bar a \mu} &=& \varkappa \left\langle Q \left|
		\hat T_{\bar a \mu}
	\right| Q \right\rangle ,
\label{2-30}\\
	\left\langle Q \left|
		\nabla_\mu \hat T_{\bar a}^{\phantom a \mu}
	\right| Q \right\rangle &=& 0,
\label{2-40}
\end{eqnarray}
where $\left\langle Q \left| \cdots \right| Q \right\rangle$ is the
quantum averaging with respect to the quantum state $\left. \left. \right| Q \right\rangle$.

Also,
to check the consistency of the ans\"atze for a 2-point Green's function,
instead of solving the Dirac equation~\eqref{2-20},
we will use the Bianchi identities  \eqref{2-40}, as we did in Ref.~\cite{Dzhunushaliev:2014daa}.

\section{Approximate model of the nonperturbative spinor vacuum}

In order to write down the energy-momentum tensor for a vacuum of a spinor field interacting with a GW,
we must have a model of a nonperturbative vacuum of the spinor field.
Let us emphasize once more that we cannot use a perturbative model of a spinor
field since in the presence of a gravitational field the set of equations \eqref{2-10} and \eqref{2-20} is a strongly nonlinear system.

Taking this into account and
following Ref.~\cite{Dzhunushaliev:2014daa},
our strategy in the formulation of
the model of the nonperturbative spinor vacuum is as follows:
(i) we write some classical ans\"atze for a spinor; (ii) we derive the corresponding energy-momentum tensor; and (iii) we then write hats over the corresponding spinor components.

As the first step,  we take the ans\"atze for the classical spinor in the form
\begin{equation}
	\psi = e^{i \omega \eta} \begin{pmatrix}
  		A \\
  		B \\
  		V \\
  		Q \\
  	\end{pmatrix}.
\label{2-50}
\end{equation}
For the tetrad \eqref{2-75} below, there is only one nonvanishing component
$G_{\bar 1 \eta}$. Consequently, we have to choose the components of the spinor $A,B,V,Q$ so that we obtain
the corresponding component $T_{\bar 1 \eta}$. This happens if $B=A$ and $Q=V$. In this case
we have
\begin{equation}
	T_{\bar 1 \eta} =
	- 2 V V^*
  \left(
		\beta' \cosh \beta + \alpha' \sinh \beta + 4 \omega
	\right),
\label{2-60}
\end{equation}
where the prime denotes $d/d\eta$.

With the ans\"atze \eqref{2-50} and the energy-momentum tensor \eqref{2-60} in hand,
we suggest the following approximate model of the nonperturbative vacuum of the spinor field.
\begin{itemize}
  \item The nonperturbative vacuum is described by the following operator of the spinor field:
      \begin{equation}
	       \hat \psi = e^{i \omega \eta} \begin{pmatrix}
  		    \hat A \\
  		    \hat A \\
  		    \hat V \\
  		    \hat V \\
  	  \end{pmatrix}.
      \label{2-70}
      \end{equation}
      The constant operators $\hat A, \hat V$ appearing here are independent of $\eta$.
  \item The corresponding energy-momentum tensor of the spinor field is
  \begin{equation}
	 \left\langle Q \left|\hat T_{\bar 1 \eta} \right| Q \right\rangle =
	 - 2 \left\langle \hat V \hat V^\dagger \right\rangle
    \left(
		  \beta' \cosh \beta + \alpha' \sinh \beta + 4 \omega
	 \right).
  \label{2-71}
  \end{equation}
  \item To check this model, we calculate the divergence of the energy-momentum tensor and show that it vanishes.
\end{itemize}

\section{Einstein equations for a strong GW interacting with the nonperturbative spinor vacuum}

For the metric \eqref{1-10},
we seek a solution of the Einstein equations for a strong GW
propagating through the nonperturbative spinor vacuum. The tetrad for this metric is
\begin{equation}
	e^{\bar a}_{\phantom a \mu} =
	\begin{pmatrix}
  		0 & 1 	&	0	&	0 \\
  		1 & 0	&	0	&	0 \\
  		0	&	0	&	
  		\chi(\eta) e^{\alpha(\eta)/2} &	0 \\
  		0	&	0	&	\chi(\eta) e^{\alpha(\eta)/2} \sinh \beta(\eta)	&	
  \chi(\eta) e^{-\alpha(\eta)/2}
 	\end{pmatrix}
\label{2-75}
\end{equation}
with the corresponding two-dimensional metric
\begin{equation}
	\gamma_{ab} =
	\begin{pmatrix}
  		e^{\alpha} \cosh^2 \beta  & \sinh \beta 	 \\
  		\sinh \beta               & e^{-\alpha}	 \\
 	\end{pmatrix} .
\label{2-77}
\end{equation}
Here $\alpha(\eta)$ and $\beta(\eta)$ are arbitrary functions~\cite{lanlif}.

The substitution of the tetrad \eqref{2-75} and the energy-momentum tensor \eqref{2-71} into the Einstein equations \eqref{2-30} yields the equation
\begin{equation}
	- \chi^{\prime \prime} +  \left [
    \beta' \cosh \beta + \alpha' \sinh \beta -
		\frac{1}{4} \left({{\alpha^\prime}^2+\beta^\prime}^2\right)\cosh^2 \beta-
    \frac{1}{4} \alpha^\prime \beta^\prime \sinh 2 \beta
	\right ]\chi = -4\tilde \omega \chi,
\label{2-80}
\end{equation}
where the prime denotes differentiation with respect to the dimensionless
$
\tilde \eta = \varkappa \left\langle \hat V \hat V^\dagger \right\rangle \eta
$,
and the dimensionless ${\tilde \omega =\omega/\left(\varkappa \left\langle \hat V \hat V^\dagger \right\rangle\right)}$.
(For convenience, we omit the tilde from $\tilde \eta$ and $\tilde \omega$ hereafter.)
The terms $\beta' \cosh \beta, \alpha' \sinh \beta$ on the left-hand side
and $4\omega$ on the right-hand side of Eq.~\eqref{2-80} are the imprints of the nonperturbative spinor vacuum.

One sees immediately that Eq.~\eqref{2-80} is a Schr\"odinger-like equation with the effective potential
\begin{equation}
	V_{eff} =
    \beta' \cosh \beta + \alpha' \sinh \beta -
		\frac{1}{4} \left({{\alpha^\prime}^2+\beta^\prime}^2\right)\cosh^2 \beta-
    \frac{1}{4} \alpha^\prime \beta^\prime \sinh 2 \beta.
\label{2-90}
\end{equation}

In what follows
we will seek periodic solutions to Eq.~\eqref{2-80}. It is clear that for periodic functions $\alpha, \beta$
we obtain an equation similar to that describing
%we have the same equation as for
the movement of a single electron in a crystal. Such an equation has been well studied in the literature (see, for example, the textbook \cite{Harrison}),
and we can apply all mathematical methods used in solving the Schr\"odinger equation to our case of a strong GW.

\section{General solution}

In order to solve Eq.~\eqref{2-80} for a periodic potential,
in this section we apply the methods developed in solid state theory.
The only difference is that the function $\chi(\eta)$ is a real function, unlike the usual
quantum mechanics where a wave function is complex.

First of all, let us rewrite Eq.~\eqref{2-80} in the form
\begin{equation}
	- \chi^{\prime \prime}(\eta) +  V_{eff} (\eta) \chi (\eta)=
  -4 \omega \chi (\eta).
\label{8-10}
\end{equation}
In order to find a solution of the type of
GW, we have to investigate the case of a periodic potential %$V_{eff}$:
\begin{equation}
	V_{eff} (\eta +\eta_0) = V_{eff} (\eta),
\label{8-20}
\end{equation}
where $\eta_0$ is the period of the effective potential.

We seek a solution to Eq.~\eqref{8-10}  in the form
\begin{equation}
	\chi (\eta) = \chi_0 + \sum \limits_{k=1}^\infty \left[
    a_k \cos \left( \frac{2 \pi k}{\eta_0} \eta \right) +
    b_k \sin \left( \frac{2 \pi k}{\eta_0} \eta \right)
  \right].
\label{8-30}
\end{equation}
In order to obtain a set of equations for the coefficients $\chi_0, a_k, b_k$, we: (a) integrate Eq.~\eqref{8-10}; (b) multiply \eqref{8-10} by
$\frac{2}{\eta_0} \cos \left( \frac{2 \pi n}{\eta_0} \eta \right)$ and integrate; and (c) multiply \eqref{8-10} by
$\frac{2}{\eta_0} \sin \left( \frac{2 \pi n}{\eta_0} \eta \right)$ and integrate. This yields the following system of equations:
\begin{eqnarray}
  V_0 +
	\frac{1}{2}\sum \limits_{k=1}^\infty \left[
    a_k \tilde V_k +
    b_k \bar V_k
  \right] &=& -4 \omega \chi_0 ,
\label{8-40}\\
   \left(
    \frac{2 \pi n}{\eta_0}
  \right)^2	a_n +\chi_0\tilde V_n +
  \sum \limits_{k=1}^\infty \left[
    a_k \tilde V_{nk} +
    b_k \bar V_{nk}
  \right] &=& -4 \omega a_n ,
\label{8-50}\\
 \left(
    \frac{2 \pi n}{\eta_0}
  \right)^2	b_n +\chi_0\bar V_n +
  \sum \limits_{k=1}^\infty \left[
    a_k \bar V_{kn} +
    b_k \tilde{\tilde V}_{nk}
  \right] &=& -4 \omega b_n ,
\label{8-60}
\end{eqnarray}
where $n = 1,2, \cdots , \infty$ and
\begin{eqnarray}
	V_0  &=& \frac{\chi_0}{\eta_0}
  \int\limits_{-\eta_0/2}^{+\eta_0/2}V_{eff} d\eta;
\label{8-65}\\
	\tilde V_k  &=& \frac{2}{\eta_0}
  \int\limits_{-\eta_0/2}^{+\eta_0/2} V_{eff}
  \cos \left( \frac{2 \pi k}{\eta_0} \eta \right) d \eta,
  \quad k = 1,2, \cdots , \infty ;
\label{8-70}\\
  \bar V_k  &=& \frac{2}{\eta_0}
  \int\limits_{-\eta_0/2}^{+\eta_0/2} V_{eff}
  \sin \left( \frac{2 \pi k}{\eta_0} \eta \right) d \eta,
  \quad k = 1,2, \cdots , \infty ;
\label{8-80}\\
  \tilde V_{nk}  &=& \frac{2}{\eta_0}
  \int\limits_{-\eta_0/2}^{+\eta_0/2}
  \cos \left( \frac{2 \pi n}{\eta_0} \eta \right)
  V_{eff}
  \cos \left( \frac{2 \pi k}{\eta_0} \eta \right) d \eta,
  \quad n, k = 1,2, \cdots , \infty ;
\label{8-90}\\
  \bar V_{nk}  &=& \frac{2}{\eta_0}
  \int\limits_{-\eta_0/2}^{+\eta_0/2}
  \cos \left( \frac{2 \pi n}{\eta_0} \eta \right)
  V_{eff}
  \sin \left( \frac{2 \pi k}{\eta_0} \eta \right) d \eta,
  \quad n, k = 1,2, \cdots , \infty ;
\label{8-100}\\
  \tilde{\tilde V}_{nk}  &=& \frac{2}{\eta_0}
  \int\limits_{-\eta_0/2}^{+\eta_0/2}
  \sin \left( \frac{2 \pi n}{\eta_0} \eta \right)
  V_{eff}
  \sin \left( \frac{2 \pi k}{\eta_0} \eta \right) d \eta
  \quad n, k = 1,2, \cdots , \infty .
\label{8-110}
\end{eqnarray}

In principle, by using the set of equations \eqref{8-40}-\eqref{8-60}, one can find a GW solution for any periodic metric functions $\alpha (\eta)$ and $\beta(\eta)$.
Remarkably, in doing so, one can employ all well-developed methods of solid state theory (see, for example, the textbook~\cite{Harrison}).

\section{The Kronig-Penney model for gravitational waves}

In the previous section we have used the well-known methods of solid state theory to obtain the general solution for the GW.
To discuss some particular features of the GW, here we employ an approximate approach of solid state theory (the Kronig-Penney model)
for the description of the GW.
%That is, we want to apply the Kronig-Penney approximation a GW and discuss the features of the resulting GW. %of this type.

Let us consider the simplest case, $\beta = 0$, for which Eq.~\eqref{8-10} yields
\begin{equation}
	- \chi^{\prime \prime} -
  \frac{1}{4} {\alpha^\prime}^2 \chi = -4 \omega \chi .
\label{6-10}
\end{equation}
For such a case we have the following two-dimensional metric \eqref{2-77}:
\begin{equation}
	\gamma_{ab} =
	\begin{pmatrix}
  		e^{\alpha}   & 0	 \\
  		0            & e^{-\alpha}	 \\
 	\end{pmatrix} .
\label{6-20}
\end{equation}
In order to apply the Kronig-Penney approximation, we will use the metric function $\alpha(\eta)$ in the range $-a < \eta < b$:
\begin{equation}
	\alpha(\eta) = \begin{cases}
    C_0 &
    \text{if} \quad -a < \eta < 0, \\
    V_0 \eta + C_1 & \text{if} \quad 0 < \eta < \frac{b}{2}, \\
    - V_0 \eta + C_2 & \text{if} \quad \frac{b}{2} < \eta < b,
  \end{cases}
\label{6-30}
\end{equation}
where $C_{0,1,2}$ and $V_0$ are constants; $a+b = \eta_0$, and
\begin{equation}
	\alpha(\eta + n \eta_0) = \alpha(\eta),
\quad n=\pm 1, \pm 2, \cdots.
\label{6-40}
\end{equation}

\begin{figure}[t]
\centering
  \includegraphics[height=4cm]{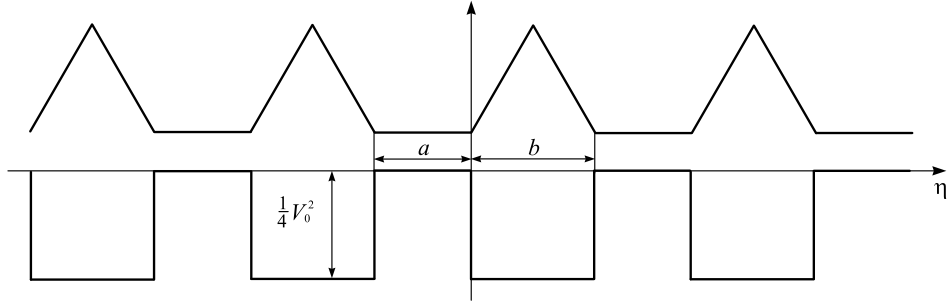}
%\vspace{-1.cm}
\caption{
The profiles of $\alpha(\eta)$ (the top graph) and
$V_{eff} = -\frac{{\alpha^\prime}^2(\eta)}{4}$ (the bottom graph).
}
\label{wave_pot_fig}
\end{figure}

The corresponding effective potential
$V_{eff} = - {\alpha^\prime}^2/4$ is
\begin{equation}
	V_{eff} = \begin{cases}
    0 &
    \text{if} \quad -a < \eta < 0, \\
    -\frac{V_0^2}{4} & \text{if} \quad 0 < \eta < b
  \end{cases}
\label{6-50}
\end{equation}
with the periodicity
\begin{equation}
	V_{eff}(\eta + n \eta_0) = V_{eff}(\eta), \quad n=\pm 1, \pm 2, \cdots.
\label{6-60}
\end{equation}
The profiles of $\alpha$ and $V_{eff}$ are shown in Fig.~\ref{wave_pot_fig}.

To solve Eq.~\eqref{6-10} with the effective potential \eqref{6-50}, we will employ the Kronig-Penney method
exactly in the same way
as in solid state theory \emph{but remembering that we have to seek a real solution instead of a complex one, as is done for the Schr\"odinger equation}.

For the periodic potential \eqref{6-50}, Eq.~\eqref{6-10} takes the form

\begin{equation}
\begin{cases}
    \chi^{\prime \prime} = K^2 \chi  &
    \text{if} \quad -a < \eta < 0, \\
  \chi^{\prime \prime} = - Q^2 \chi &
    \text{if} \quad 0 < \eta < b,
  \end{cases}
\label{6-80}
\end{equation}
where $K^2 =4 \omega$, $Q^2 = V_0^2/4 -K^2$.

For the fundamental region $-a \leq \eta \leq b$, we seek a solution in the following form
\begin{equation}
	\chi(\eta) = \begin{cases}
    f e^{K \eta} + g e^{-K \eta} &
    \text{if} \quad -a < \eta < 0, \\
    c \sin Q \eta + d \cos Q \eta &
    \text{if} \quad 0 < \eta < b .
  \end{cases}
\label{6-90}
\end{equation}
Using these,
we first constrain continuity inside the fundamental domain (i.e., for $\eta=0$):
\begin{equation}
\begin{cases}
	f + g = d ,
\\
  K(f - g) = Q c .
  \end{cases}
\label{6-100}
\end{equation}
For the region $b \leq \eta \leq a+b$, we seek a solution in the form
\begin{equation}
	\chi(\eta) = f e^{K [\eta-(a+b)]} + g e^{-K [\eta-(a+b)]} .
\label{6-120}
\end{equation}
We then constrain continuity outside the fundamental domain (i.e., for $\eta=b$):
\begin{equation}
\begin{cases}
	c \sin Qb + d \cos Qb = f e^{-K a} + g e^{K a},
\\
  c\, Q \cos Qb - d Q \sin Qb = f K e^{-K a} - g K e^{K a} .
  \end{cases}
\label{6-130}
\end{equation}

A necessary condition for the set of equations \eqref{6-100} and \eqref{6-130} to have
a nontrivial solution is that the determinant of the corresponding matrix is zero:
\begin{equation}
  \begin{vmatrix}
  	0 & 1 	&	-1	&	-1 \\
  	Q & 0	&	-K	&	K \\
  	\sin Qb	&	\cos Qb	&	-e^{-K a} &	- e^{K a} \\
  	Q \cos Qb	&	-Q \sin Qb	&	-K e^{-K a} &	K e^{K a}
 	\end{vmatrix} = 0 .
\label{6-150}
\end{equation}
The resulting constraint equation
%After some calculations we have the constraining equation
for the parameters $K,Q,a,$ and $b$ is
\begin{equation}
  \frac{K^2 - Q^2}{2 K Q} \sinh aK \sin b Q +
  \cosh aK \cos bQ = 1.
\label{6-160}
\end{equation}
Introducing the dimensionless parameters
\begin{equation}
  x = \frac{4 \omega}{V_0^2/4}, \quad
  \tilde a = \frac{a V_0}{2}, \quad
  \tilde b = \frac{b V_0}{2},
\label{6-170}
\end{equation}
we obtain from \eqref{6-160}
\begin{equation}
  \frac{2 x-1}{2 \sqrt{x-x^2}}
  \sinh \left( \tilde a \sqrt{x} \right)
  \sin \left( \tilde b \sqrt{1-x} \right) +
  \cosh \left( \tilde a \sqrt{x} \right)
  \cos \left( \tilde b \sqrt{1-x} \right) = 1.
\label{6-180}
\end{equation}
Here $0<x\leq1$, i.e., $0<\omega\leq V_0^2/16$.
The latter expression is obtained from the condition  $Q^2\geq 0$
which has been assumed to be satisfied in the derivation of the above results.
The case of $Q^2<0$ is discussed below.

\begin{figure}[t]
\centering
		\includegraphics[height=9cm]{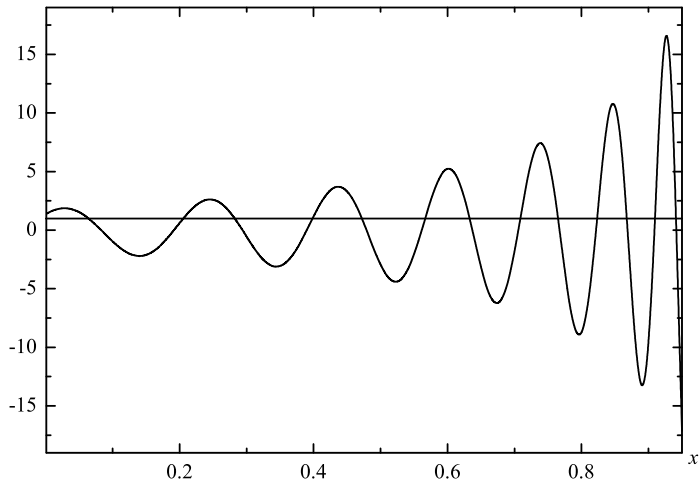}
\vspace{-1.cm}
	  \caption{The profile of the left-hand side of the constraint equation \eqref{6-180} for $\tilde a = 3$ and $\tilde b = 50$.
The straight line corresponds to~1.}
  \label{dispersion}
\end{figure}

The typical profile of the left-hand side (lhs) of the constraint equation \eqref{6-180} is shown in Fig.~\ref{dispersion}.
One can see that there is a finite number $n$ of possible values of $\omega_n$ which depends on the values of the parameters $\tilde a$ and $\tilde b$.
The numerical values of $\omega_n$ are determined by the points of intersection of the lines in Fig.~\ref{dispersion}.
Note that in the limit $x\to 0$ the  $\text{lhs}=-(\tilde a/2) \sin \tilde b+\cos \tilde b$, and when $x=1$, the  $\text{lhs}=(\tilde b/2) \sinh \tilde a+\cosh \tilde a$,
i.e., the lhs remains always finite.

The case of $Q^2 < 0$ can be obtained from \eqref{6-160} by changing
$Q$ to $i Q$:
\begin{equation}
  \frac{K^2 + Q^2}{2 K Q} \sinh aK \sinh b Q +
  \cosh aK \cosh bQ = 1.
\label{6-190}
\end{equation}
The corresponding dimensionless equation will then be
\begin{equation}
 \frac{2x-1}{2 \sqrt{x^2-x}}
  \sinh \left( \tilde a \sqrt{x} \right)
  \sinh \left( \tilde b \sqrt{x-1} \right) +
  \cosh \left( \tilde a \sqrt{x} \right)
  \cosh \left( \tilde b \sqrt{x-1} \right) = 1.
\label{6-200}
\end{equation}
Here $x>1$, i.e., $\omega> V_0^2/16$.
One can easily see that the lhs of this expression is always greater than 1, i.e.,
Eq.~\eqref{6-200} does not have a solution. Physically, this means that there are no GWs with $\omega > V_0^2/16$.

\section{Experimental verification}

Let us discuss now possible experimental consequences coming from a consideration of the propagation
of a strong GW through the nonperturbative spinor vacuum.

Direct calculations show that for such a case
%for the model of a nonperturbative spinor vacuum considered here,
there exists the following electric current
\begin{equation}
  j^{\mu} = \bar \psi \gamma^{\mu} \psi =
  \left(
        4 \left\langle Q \left| \hat V \hat V^\dagger \right| Q \right\rangle , 0,
    0, 0
  \right),
\label{7-10}
\end{equation}
where $\mu = \xi ,\eta, y, z$.

%Using our definition of $\xi, \eta$ coordinates
Making the transformation to Cartesian coordinates $t,x$ as
\begin{equation}
  \xi = \frac{t - x}{\sqrt 2}, \quad
  \eta = \frac{t + x}{\sqrt 2},
\label{7-15}
\end{equation}
and similarly,
 %the same for $j^\mu$
\begin{equation}
  j^\xi = \frac{j^t - j^x}{\sqrt 2}, \quad
  j^\eta = \frac{j^t + j^x}{\sqrt 2},
\label{7-17}
\end{equation}
we have
\begin{equation}
  j^t = -j^x = 2 \sqrt{2} \left\langle Q \left|
    \hat V \hat V^\dagger
  \right| Q \right\rangle.
\label{7-20}
\end{equation}
This means that we have the electric charge and current densities concomitant with the GW.
The electric current is directed along the direction of propagation of the GW.
This observation allows us to suggest the following experimental verification
of the GW and nonperturbative spinor vacuum models
considered here: \textcolor{blue}{\emph{It is possible to try to measure
the weak electric charge and current densities together with the standard measurements of GWs (LIGO, LISA, and so on).}} On the other hand,
instead of measurements of the electric current, one can measure the corresponding magnetic field.

\section{The correctness of  the ans\"atze for the spinor field
%The correctness of the model of a nonperturbative spinor vacuum
}

The correctness of Eq.~\eqref{2-40} is verified directly. To do this,
one can calculate the expression $\nabla_\mu \hat T_{\bar a}^{\phantom a \mu}$ as the classical one and then substitute in it the 2-point Green's function
$\left\langle \hat \psi^* \hat \psi \right\rangle$.
Then, taking into account that we have only one nonvanishing component of the energy-momentum tenor
$T_{\bar 1 \eta}$ [see Eq.~\eqref{2-60}], the Bianchi identities~\eqref{2-40} are trivially satisfied.

\section{Conclusion}

To summarize our results:
\begin{itemize}
  \item We have suggested an approximate model of the nonperturbative spinor vacuum.
  \item The propagation of strong gravitational waves interacting with such a vacuum has been investigated.
  \item It was shown that the Einstein equation reduces to a Schr\"odinger-like equation with a periodic potential.
  %  describing an electron moving in a periodic potential.
  \item We have obtained the solution for the special case of the metric function corresponding to a GW with the $g_{yy,zz}$ components only.
  \item A possible experimental verification of the nonperturbative spinor vacuum model interacting with a strong gravitational wave has been suggested.
\end{itemize}

For better qualitative understanding of some features of the resulting
gravitational waves, we have considered the simplest case where the periodic potential is modeled
by the Kronig-Penney potential of solid state theory. As a result, it was shown that the GW parameters (the wave length $\eta_0$ and the amplitude $V_0$)
can be arbitrary but the parameter $\omega_n$, affecting the form of the plane wave in the spinor vacuum,  is quantized.

The correctness of the approximate model of the nonperturbative spinor vacuum was verified by
calculating the divergence of the right-hand side of the Einstein equations. One might expect that for an improved model
of the nonperturbative vacuum it will be necessary to use the nonperturbatively quantized Dirac equation where the term ``spin connection $\times$ spinor field''
(i.e., $\left\langle Q \left| \omega_{ab \mu} \psi
\right| Q \right\rangle$)
need to be taken into account. The latter means that we have to take into account the quantum correlation between a metric and a spinor field.

Lastly, it may be noted that the formal analogy of the GW equation here obtained with the Schr\"odinger equation
%the equation for a GW here obtained is formally identical with the Schr\"odinger equation. This
allows the possibility of using the well-developed methods from other fields of physics
to solve such an equation. In particular, for the periodic metric functions discussed here, Eq.~\eqref{8-10} is identical in form to the equation of motion
of an electron in a crystal. Aside from the formal aspect, such an analogy would lead us to use physical intuition to obtain further results in this area,
as it is done, for example, in analogue gravity~\cite{Barcelo:2005fc}.

\section*{Acknowledgments}

This work was partially supported  by a grant in fundamental research in natural sciences by the Science Committee of the Ministry of Education and Science of Kazakhstan and by a grant of the Volkswagen Foundation.

\end{document}